\documentclass[11pt,twoside]{article}

\usepackage{asp2006}
\usepackage{epsf}
\usepackage{lscape}

\begin{document}
\title{SOME OBSERVATIONAL ASPECTS OF R CORONAE BOREALIS STARS}   
\author{N. Kameswara Rao}   
\affil{Indian Institute of Astrophysics; Bangalore,  560034 India}    

\begin{abstract} 

Some of the observational aspects related to the
   evolutionary status and dust production in  R Cor Bor stars are discussed.
    Recent work regarding the  surface abundances, stellar winds and
    evidence for  dust production in these high luminosty hydrogen
   deficient stars are also reviewed. Possibility of the stellar winds
    being maintained by surface magnetic fields is also considered.

\end{abstract}


\section{Introduction}   

R Coronae Borealis stars (RCBs) are dust making hydrogen
  deficient, carbon and helium rich F-G supergiants. The major puzzles they
  pose are two: namely, what are their orgins (i.e., their place in the stellar
  evolution of single or double stars, their  predecessors and
  descendants) and how do they make dust in their warm atmospheres.
  I would like to explore in this article some of the
  observational aspects that concern these two issues. There are two major
  scenarios that have been proposed to account for their orgins. The first
  one involves a single star -- a result of a `final He-shell flash' in a
   post-asymptotic giant branch
  star that is passing through  the white dwarf cooling track. The energy
  released is expected to make the star a hydrogen deficient, cool luminous
  star that evolves relatively rapidly across the HR diagram back to the
  white dwarf track for a second time. This is dubbed as `final flash (FF)'
  scenario. The second one involves a pair of white dwarfs, a C-O white dwarf
  and a He white dwarf that merge togather by emitting gravitational waves.
  The He white dwarf is accreted on to the C-O white dwarf leading to the
  formation of H-poor cool supergiant with a white dwarf at its core. This is
  dubbed as `double-degenerate (DD)' scenario. The life times in H-poor
  supergiant stage are supposed to be longer in the DD scenario. Although there
  are no decisive observational tests, presently, to choose either one
  (or any other alternative) of the scenarios, the accumalted observations so far seem
  to favour the DD scenario.

\section{The Number and Distribution of RCBs}

Do the estimated number of RCB stars in the galaxy provide any
  clues regarding the two scenarios proposed for their orgins?
  Presently known members in the galaxy amount
  to about 40 \citep{zanie2005}, about 21 are known in LMC (including
  DY\,Per stars) and about 2 in SMC \citep{alcock2001,kraemer2005}.
  The statistics is more complete for LMC objects. By scaling the LMC
  population to the Galaxy, Alcock et al. estimate the total number of RCBs
  in the Galaxy as over 3200. According to \citet{iben1996} the two
   evolutionary scenarios proposed
  (DD and FF) individually  could only account partially for the
   number of RCBs present in the Galaxy. Thus both schemes might be contributing
  to the total number. However, based on the data of
  white dwarf  binaries  from SPY project, \citet{mitclay2005}
  claim that an estimate of the
   expected population
  of RCBs in the galaxy produced by DD scenario is consistent with numbers
  expected in the Galaxy. A detailed reassessment using updated evolutionary
  time scales would be worth while.

           The galactic distribution of RCBs seem to suggest that they might
  belong to the thick disk population \citep{zanie2005} although
  few of them might even be part of the halo e.g. U\,Aqr \citep{cotlaw1998}.
  The existence of RCBs in the galactic bulge as well as SMC suggets
  a spread in metallicity. The similarity of kinematics with planetary nebulae
of Peimbert's III group seem to suggest that majority of them are likely to be
 moderately metal poor \citep{raolamb1996}.

\section{Properties at Maximum light}

\subsection{Spectral analyses}

Most of the spectral analyses that were attempted to arrive
   at the surface abundances and other properties like Teff and log g etc
   are based on the LTE line blanketted model atmospheres computed at Uppsala
   (Asplund, Gustafsson, Eriksson) for $T_{\rm eff}$ $<$ 9000 K and at Armagh (Jeffery
  and collaboraters) for $T_{\rm eff}$ $>$ 9000K. Some of the major uncertainties in
  these analyses are the C/He ratio (the number density of carbon to helium)
   and the famous `carbon problem' \citep{asp2000}. The C/He is needed
 to estimate the
  mass fraction of elements and cannot directly be estimated from observations.
  It has been assumed as 1\%, same as the mean  estimated value from observations
  of extreme helium stars (EHes) (see furthur for a comment on this assumption).
   This value seems to be consistent with the metallicity expected from the
  galactic distribution \citep{raolamb1996,pan2001}.

\subsection{Carbon Problem}

The problem (which would be discussed in detail in
 Gustafsson's review) is that the carbon abundance estimated from the observed
 C\,{\sc i} lines is four times less than the input abundance of carbon (based on C/He).
  in the model. Where did the missing carbon go. The answer to this puzzle is
 still missing. The continuous opacity in the
  atmosphere of RCBs is controlled by the photoionization of C\,{\sc i} from excited
 levels and the observed lines arise from the levels marginally lower in excitation
 potential. Thus the equivalent width of C\,{\sc i} lines are independent of $T_{\rm eff}$ and
$\log g$ (and mildly dependent on microturbulence). This is confirmed by
 observations. Figure 1 shows the spectrum of six RCBs over plotted in the
 region of 6380 to 6412 \AA. 
Lines of C\,{\sc i} are of same strength in all the stars
 where as lines of other elements like Fe\,{\sc i}, Fe\,{\sc ii} vary enormously in strength.
 A similar comparision of the spectrum of RY\,Sgr ($T_{\rm eff}$ $\sim$ 7200K) and H rich,
 Fe poor RCB star V854\,Cen ($T_{\rm eff}$ $\sim$ 6750K) also shows that the C\,{\sc i} lines are
 mariginally stronger in RY\,Sgr relative to V854\,Cen in the 6380 to 6412 \AA\
 region as expected.
 However a comparision of these two stars in the UV region 1500 to 1800 \AA\
 shows a very different appearance (see the figure in Clayton \& Ayers
 2001 illustrating the HST spectra). The C\,{\sc i} lines eg. 1657 \AA\ in V854\,Cen
 is very much stronger than in RY\,Sgr contrary to that seen in optical region.
 I asked my friend Martin Asplund `what is the cause of this?'
 He did a qualitative computation of the line strenths of C\,{\sc i} lines in both
 stars using abundances obtained from the optical region \citep{asp1998,asp2000}
 and told me that C\,{\sc i} is not the source of continuous opacity in the UV region,
 in fact it is Si\,{\sc i} photoionization that provides the continuos opacity
 thus carbon becomes a trace element and the line strengths computed are similar
 to that observed. It is in principle possible to obtain the correct carbon
 abundance by analysing the C\,{\sc i} lines in UV, like the other elemental
 abundances estimated in optical region. This approach might show a way to
resolve the carbon problem.
                    In analysing the [C\,{\sc i}] lines in RCBs, \citet{pan2004}
 suggest that a chromosphere like temerature rise in the atmosphere of the stars
 might provide a solution to the carbon problem. However, inspite of the
 carbon problem, \citet{asp2000} found the abundance ratios are unaffected.

\section{Abundance Patterns}

The surface abundance patterns in RCBs and EHes have
 recently been reviewed by \citet{rao2005}. The surface abundances provide the most
 vital clues to wards the understanding the origins of these stars.
 The first major study of the surface
 abundances of large number of RCBs by \citet{raolamb94} revealed
 two distinct patterns in [Si/Fe] and [S/Fe] plot. Majority of the stars (14 out
 of 18) showed these ratios to be around 0.5 with a mild Fe deficiency relative
 to solar, where as minority of four stars showed very high values of [Si/Fe]
 and [S/Fe]  with a large deficiency of Fe. A similar pattern of majority
 and minority is also seen in EHe stars (see Rao \& Lambert 2007 for a recent plot
 of [Si/Fe] vs [S/Fe]). So far mainly the warm RCBs (F,G type) have been
 analysed spectroscopically. The cool RCB and DY\,Per stars need to be analysed
 comprehensively for abundances although few attempts have been made by 
 \citet{kip2002,kipklo2006,zac2005,zac2007}.
  The mean abundance of
 several elements in  the majority and
 minority groups in RCBs and EHes has been given in \citet{rao2005}. The dispersion
 around the mean in each group is surprisingly small e.g.  $\sim$ 0.27 dex for the
 majority group in RCB and EHes.
 Figure 2 (lower panel) illustrates the differences in mean abundance of elements in
 the majority group of RCB(mean of 15 stars) and EHe (mean of 12 stars).
 Figure 2 (upper panel) shows similar plot for minority group (4 RCBs and 2 for EHes).

               If RCBs and EHes are related objects (or one group evolving into
 the other) it is expected the abundance of most elements would be similar (same)
 except for those that might get affected in that particular stage of
 evolution. Thus the difference in abundances between mean RCBs and mean
 EHes are expected to be around zero. Figure 2 (lower panel) brings out two aspects
 clearly.
  Four elements H,N,Ne and Mg show significant differences in the
  majority group between
  RCBs and EHes. N is enhanced in RCBs where as H,Ne and Mg are
 enhanced in EHes suggesting $^{14}$N could have been alpha processed
 to $^{22}$Ne and $^{25}$Mg in EHes. Thus EHes might be a later phase to RCBs.
      There is a systematic shift of $-$0.3 dex for all elements suggesting that the
  the abundances of RCBs are systematically lower to that of EHes. This might be a
  consequence of the assumption, C/He being same (1\%) for both groups,
  since the mean metallicity of RCBs is not expected to be lower than that
  of EHes from the galactic kinematics and distribution. Thus C/He for
 the majority RCBs is suggested to be higher than that of majority EHes
 so that the metallicity of both groups remain the same.

\begin{figure}
\plottwo{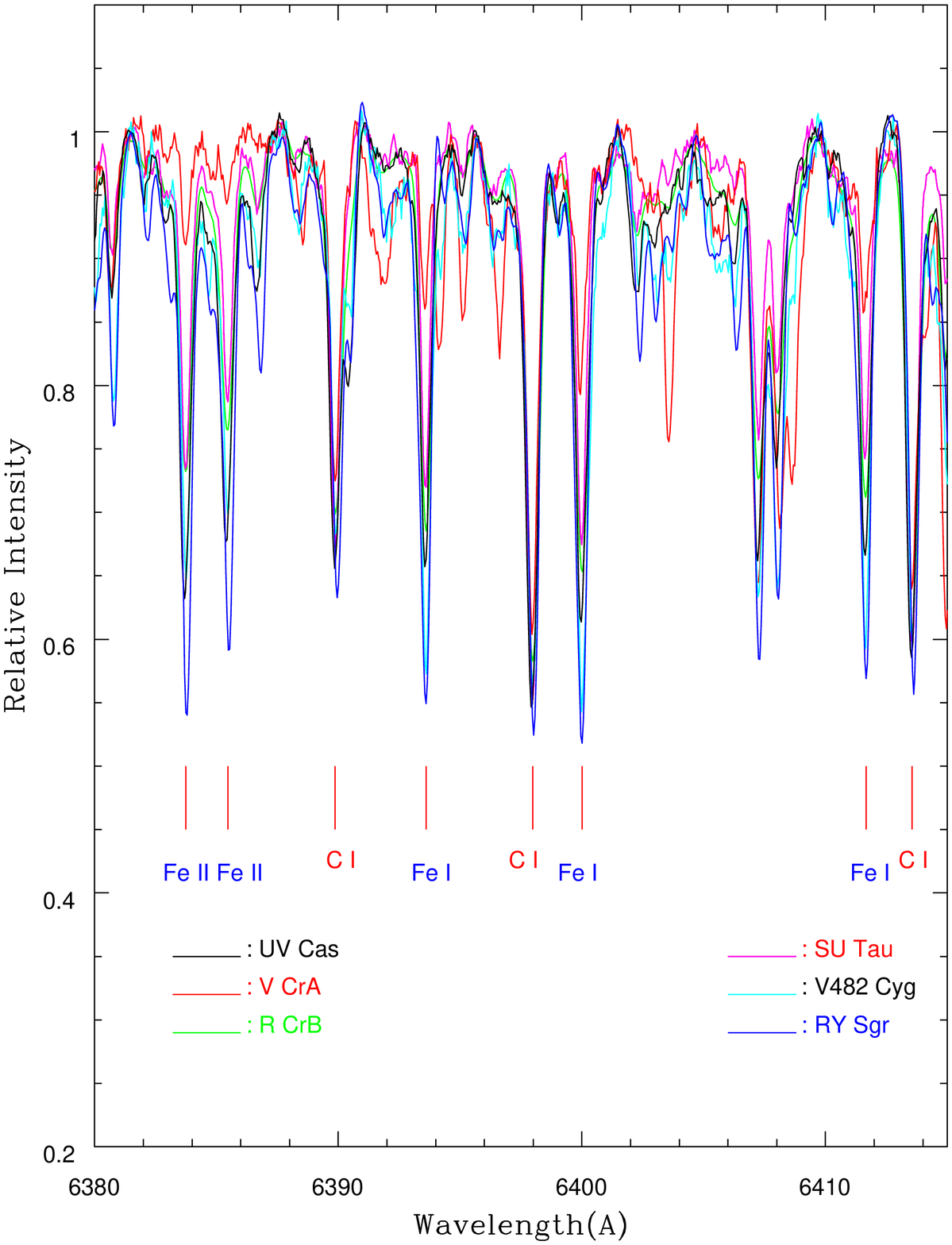}{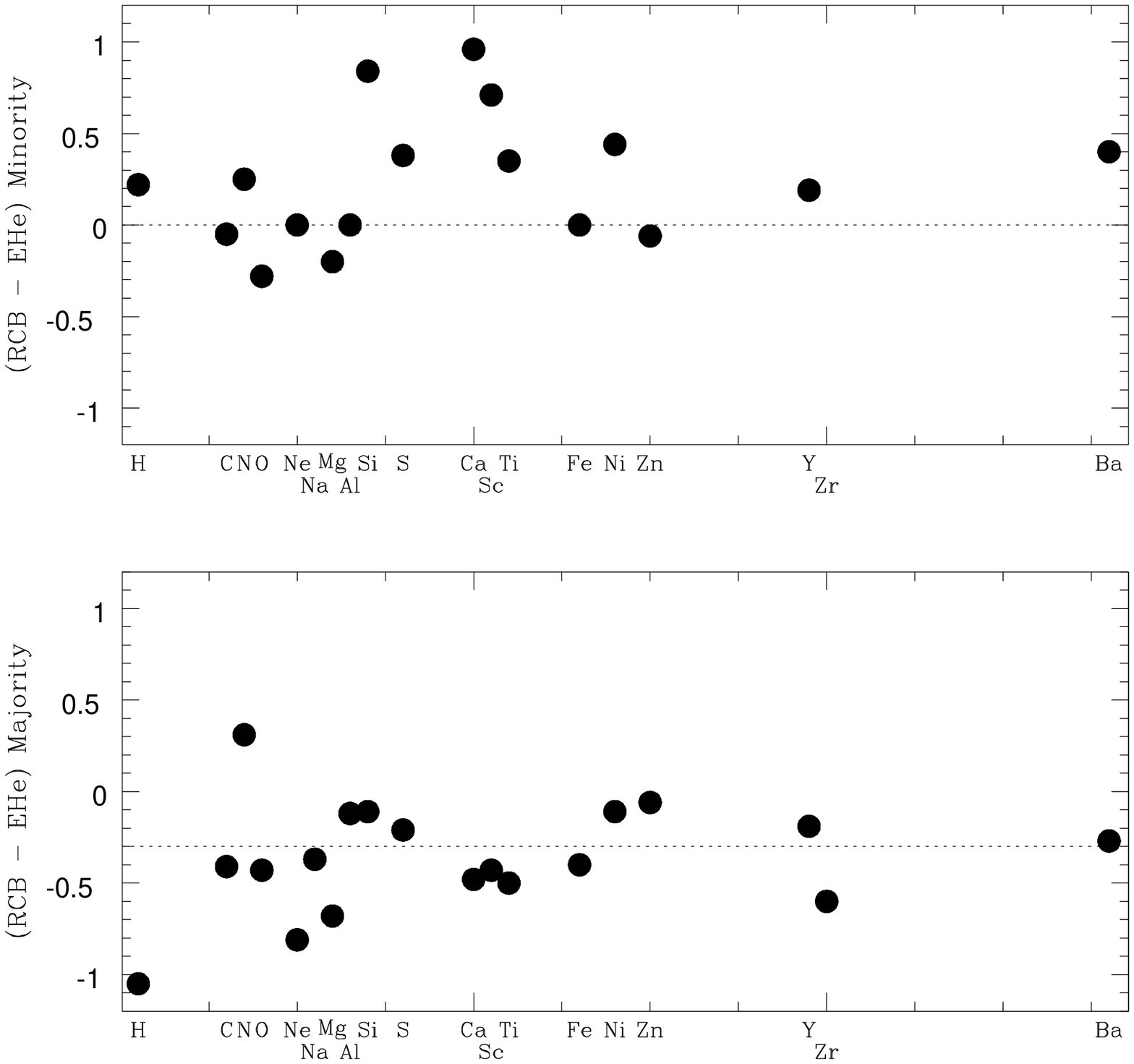}
\caption{Spectra in the region of 6380 to 6415 \AA\ of six RCB stars are
           supperposed to illustrate the constancy of C\,{\sc i} line strengths
           in these stars of varying $T_{\rm eff}$ and metallicity. Note the variation
           in strength of Fe\,{\sc i} and Fe\,{\sc ii} lines where as the variations
           in C\,{\sc i} lines are minimal. \label{fig1}}
\caption{lower panel: The difference in abundances of RCBs and EHes of majority
            group. Except for H, N, Ne and Mg the abundances in these two
            types of stars are similar but shifted by a constant factor
            possibly due to a difference in C/He ratio.
            upper panel: A similar plot as lower panel but for the minority group. Si,S, Ca, Sc
            are enhanced in RCBs. All the rest of the elements have similar
            abundances (within the errors). \label{fig2}}
\end{figure}

On the other hand in the minority group (although the numbers are small)
  Figure 2 (upper panel) shows no systematic shift for most of the elements except
   Si,S,Ca,Sc which are enhanced in RCBs. Thus the C/He ratio might be
   same for both groups. Generally the elements Si,Ca, Sc, may be S
   get locked up in grains, the enhanced abundance of these elements in
   RCBs might be a consequence of dust production in minority RCBs?

\subsection{Comments on individual elements}

   The Hydrogen abundance in majority of RCBs is low by a dex relative
   to majority EHes. The hydrogen abundance varies enormously in the
   group $<$ 4.0 to 7.0. The minority objects show even higher H abundance.
   \citet{kip2002,kipklo2006,zac2005,zac2007} have made
   estimates of H-abundance in cool RCB stars Z\,Umi, DY\,Per using the
   H$\alpha$ region in these stars. Estimates based on the presence of such a high
   excitaion line in the region with heavy blending due to CN, C$_2$ molecules
   is going to be very uncertain. A better way would be to use the G-band
   lines (CH) in the blue. Figure 3 shows the synthesis of the spectral
   region in the cool RCB star U\,Aqr, using Uppsala models and line lists
   provided by Betrand Plez. An atmospheric model of $T_{\rm eff}$=5400K and 
   $\log g$=1.0 seem to be appropriate to the star and leads to estimate of H abundance of 9.5.
   Although there are some uncertainities related to the metallicity, the
   cool RCBs (and DY\,Per) seem to indicate higher H abundance in the range
   8.5 to 9.5.
              Incidentally studies of the spectra of DY\,Per, both at maximum
   and minimum light suggest that it does belong to RCB group 
   \citep{raolambcarl2007,zac2007}.
   The question whether all cool RCB (and HdC) stars show higher H
  abundance needs to be explored.

\begin{figure}
\plotone{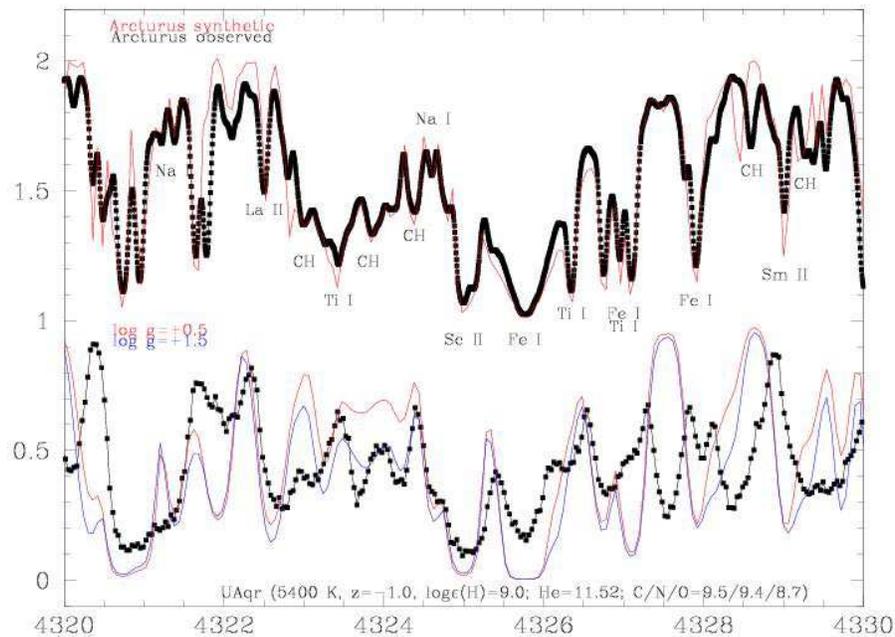}
\caption{The synthesis of the G-band region in U\,Aqr to illustrate the
           estimation of hydrogen abundance in cooler RCBs. \label{fig3}}
\end{figure}

        It is often asserted that the ratio of $^{12}$C/$^{13}$C is high in RCBs, few
 exception still exist. V\,CrA the minority RCB shows a value between 4 to 10
 \citep{raolamb2007}, and DY\,Per, which shows presence of $^{12}$C$^{13}$C bands
 in the optical spectrum, is estimated to have a value around 
 20 \citep{raolambcarl2007,zac2007}.

        The nitrogen abundance in RCBs is high suggestive of not only
 conversion of intitial C and O to N but even furthur conversion of
 synthesized C to N. The N abundance in many RCBs and EHes imply wholesale
 conversion of O to N via ON cycles. Many stars are not O deficient suggesting
 O is synthesized along with C.

        The dramatic discovery of large abundance of $^{18}$O and
   low abundance ratio of $^{16}$O/$^{18}$O in HdC stars and cool
  RCBs by \citet{clay2005,clay2007} brought DD scenario into fore. High
  resolution spectroscopy of CO bands at 2.3 microns show an estimate of
  $^{16}$O/$^{18}$O of 0.3 to 0.5 for three HdC stars and a value of 16 to cool RCB
  star S Aps \citep{garc2007}. Again the discovery of high
  Fluorine abundance, several hunderd times solar, in EHEs and RCBs
  seems furthur support to hot DD scenario \citep{pana2006,pan2007}.
  The absence of $^{18}$O and low abundance (or lack) of F in Sakurai's object
  \citep{geballe2002,pan2007}, which is a typical FF object
  does support the DD scenario as the origin for atleast majority RCBs.

       What about lithium and $s$-processed elements? Lithium is present in
 four of the fifteen analyzed majority RCBs and in one of the five HdCs
 \citep{raolamb94,raolamb1996} with an appreciable abundance
 of about log $\epsilon(\rm F)$=3. Even cool RCB star Z\,UMi shows high abundance of
 lithium \citep{gos1997,kipklo2006}. The orgin of
 this lithium is
  a special challenge to the DD scenario. Particularly  the
 presence of high abundance of F, $^{18}$O, and Li same time in some of the stars
 shows lot more is needed to develope a convincing explanation for their
 origins than the present form of the DD scenario.

      The $s$-process elemental abundances and their variation with metallicity in RCBs and EHes
 has been discussed by  \citet{rao2005} and \citet{pan2006}. The lighter
 $s$-process (ls) elements are enhanced relative to heavy $s$-process elements
 (hs) in both RCBs as well as EHes \citep{panred2006}. The variation of this ratio [ls/hs]
 with respect to metalicity ([Fe/H]) is quite different to that displayed
 by post-AGB objects \citep{rao2005} which follow Busso et al's (2001) model
 ST/1.5. This might suggest that $s$-processing in RCBs is not a result of
 conventional third dredgeup but could have been a result of second
 passage through
 the AGB stage. Although lot of progress is achieved in the study of
 surface abundances of RCBs, clarity about their origins is yet to emerge.

\section{Stellar Winds}

     \citet{clay2003} discovered that the He\,{\sc i} 10830 \AA\ line in many of the
 RCBs show P-cygni type or blue shifted strong absorption profiles indicative
 of shell of hot gas moving with  velocities over 200 km s$^{-1}$, much above
 their escape velocity.
 Several cool RCBs also seem to possess such P-Cyg profiles. The source of
 excitation of this hot gas and its connection with photosphere of the stars is
 not clear. \citet{rao2006} showed that in R\,CrB the strong photospheric lines,
 particularly the O\,{\sc i} 7774 \AA\ line profiles had a pronounced blue wing suggesting
 mass loss with an expansion velocity of 130 km s$^{-1}$ (predicted escape velocity
 is about 30 - 70 km s$^{-1}$). The O\,{\sc i} lines provide the
 link between the hot gas responsible for the He\,{\sc i} line and the stellar
 photosphere, a strong stellar wind. In addition to the O\,{\sc i} lines,
 other strong (or resonance) lines of
 lower excitation potential (e.g., Al\,{\sc i} 3944 \AA) also showed asymmetric profile
 and mass loss but
 with lower velocity of expansion. More over the blue wing of the low excitation
 lines show considerable variation in strength where as the high excitation
 O\,{\sc i} lines remained constant in the same period. \citet{rao2006} showed that the
 inferred wind velocity decreases with excitation potential, linking He\,{\sc i} lines
 showing high velocity part and low excitation Al\,{\sc i} lines with low velocity
 variable component of the wind. An interpretation of the extended blue wings
 is that the wind begins at the top of the photosphere and increases in velocity
 and excitation with height above its base. Such wind profiles to O\,{\sc i} 7774 \AA\
 lines have now been observed for several RCBs showing that stellar wind
 is a common
 charecteristic feature of the RCB stars (Rao \& Lambert 2007, and
 in preparation). The wind profiles in R\,CrB as well as in V\,CrA were unchanged
 even during most of the deep light minimum (particularly the wings) suggesting
stellar wind is a global phenomenon where as dust obstruction (production)
 might even be a local phenomenon effecting limited part of the star.

      It should be noted that He\,{\sc i} lines seen at light minimum of RCBs might
 represent the stellar wind profiles that are present all the time and
 unconnected to the minimum (and dust production). Thus the morphology of these
 profiles differ from that of broad Na\,{\sc i} D lines, Ca\,{\sc ii} H \& K lines, 
 K\,{\sc i} lines seen at light minimum \citep{rao99,raolamb2003,rao2006}.

        What drives the wind and provides the energy to heat the gas to
 cause He\,{\sc i} excitation? Heating of the wind may be by deposition of mechanical
 energy (sound or hydromagnetic waves). The photospheric absorption lines 
have widths indicating mass motions with velocities exceeding local sound speed.
 The Ca\,{\sc ii} infrared triplet lines (e.g., 8542 \AA) in V\,CrA and some other RCBs
 show similarity to
 solar chromospheric lines with an absorption core flanked by emission peaks.
 This may be indicative of magnetic activity. In this context we examined
 the absorption spectrum of R\,CrB in the region of 6170$-$6180 \AA\ where
 magnetically sensitive line of Fe\,{\sc i} 6173 \AA\ occurs. Figure 4
 shows the region in R\,CrB on several occassions. The Fe\,{\sc i} lines 6173.34
 and 6180.22 \AA\ have very similar lower excitation potentials (2.248 and 2.759 eV)
 and similar log $gf$ values ($-$2.88, $-$2.78) but differ in their Lande g factor
 (2.50 and 0.600), respectively. The magnetically sensitive  6173.3 \AA\ line with
  higher Lande g value
 shows large variations when comared to 6180.2 \AA\ line that is magnetically much
  less sensitive. May be magnetic fields play a role. This is an aspect that
  needs to be studied.

\begin{figure}
\plotfiddle{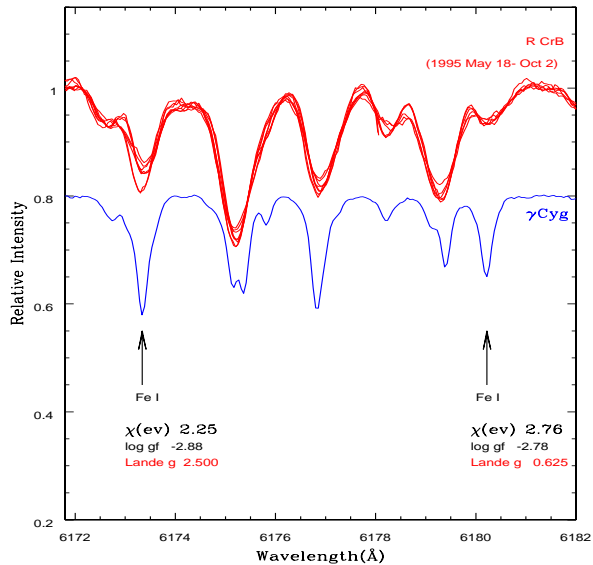}{3.0in}{0}{40}{30}{-140}{2}
\caption{The several spectra of R\,CrB at maximum light are supperposed
           to illustrate the large variation in the magnetically sensitive
           Fe\,{\sc i} line at 6173.3 \AA\ where as the less sensitive line of similar
           excitation and log gf value at 6180.2 \AA\ remained more or less constant
           in strength. \label{fig4}}
\end{figure}

\section{Pulsations and Dust production}

It is well known that RCBs show a semi-periodic variations attributed
 to pulsations \citep{lawcot1997,lawkil1996}.
 \citet{pug1977} was the
 first to make an attempt to link the deep light declines to the pulsation
 phase of
 the stars. So far this linkup could not be convincingly demonstrated for any
 RCB star. Recently \citet{cra2007} in a systematic study of four RCBs over
 several decade of photometry been able to show that pulsation is linked to
 the deep declines. Although decline does not happen at every pulsation cycle
 but when a deep decline occurs it occurs at a specific phase of pulsation
 cycle. This for the first time links the pulsation to dust production.

        Pulsation linked models of dust production have been proposed by
 \citet{woi1996} for RCBs. In this model pulsation induced shocks propogate
 through the atmosphere and when shock amplitudes get stronger the gas
  behind the shock front (preshocked gas) could attain temperatures and
 densities sufficiently
 cool and dense for the nucleation to occur. It is of utmost intrest to see
 whether such a cool gas can be detected during such decline events.
 Such cool gas has been detected in three RCBs (V854\,Cen, R\,CrB and V\,CrA)
 during the light minima through the observations of absorption lines of
 C$_2$ molecule
 arising from the Phillips system (i.e the ground electronic state of the
 molecule). The rotational temperature of 1100 K to 1230 K have been deduced
 for the gas \citep{raolamb2000,raolamb2007,rao2006} which is well
 with in the range of temperatures expected for nucleation of carbon dust.
 This detection certainly provides confidence that dust condensation does
 indeed takes place during these large light decline events. It is required
to study the atmospheric phenomenon in detail during the passage of the shock
 and link the observations to the models.

\section{RCBs and Helium enrichment in globular clusters}

I would like to conclude the observational aspects with a suggestion.  
It is now becoming very clear that multiple main sequences and extended blue horizontal branches exist in
 massive globular clusters like omega Cen, NGC\,2808 etc. These sequences can
 only be explained by a sequential enrichment of He in intercluster gas and
 subsequent star formation \citep{pio2005,pio2007}. It is suggested that
 multiple stellar generations and He enrichment to the extent of Y of 0.4 (mass
 fraction) could only take place in clusters which have deep potential wells
 favouring the retention of low-velocity stellar winds. Various species that
 could creat He enriched winds have been investigated ranging from Suprnovae to
 AGB stars. Most promising source seemed to be slow winds from AGB stars.
 Even this source is not apparently sufficient to enrich the He content to the
 required degree \citep{kara2006}. In addition, the large mass fraction
 of the He-enriched population relative to the enriching one would apparently
 `require an extremely flat initial mass function' \citep{pio2007}.
  In this context it is significant to note the discovery of ZNG1 in the
 globular cluster M5 as a hot hydrogen deficient carbon rich post-AGB star by
 \citet{dix2004}. M5 is also a cluster having extended horizontal branch with a possible gap.
  The surface properties and abundances obtained by Dixon et al. for ZNG1 are
 very similar to EHes and RCBs including C/He ratio (2\%). Thus, it is expected
 that there would be a population of hydrogen deficient stars in the globular clusters.
 If DD scenario (double degenerate binaries) is prefered for their 
 orgins, (as the current thinking suggests) then their life times as HdC
 luminous mass loosing stars would be longer and the mass of the ejecta would
 also be larger (than for single stars). They would provide mostly helium gas
without enriching too much carbon (C/He of 1\%). The final flash objects
 are expected to show comparble carbon (and oxgen) to helium (C/He of 40-50\%)
that might pose problems.
 Quantitative estimates for the population of HdC stars in globlar clusters is
 lacking, their importance to the helium enrichment might be significant. At the
 least they might compliment the helium enrichment to the AGB star ejecta
 to the intercluster medium. It would be worth while to evaluate this aspect
 in massive globular clusters as the helium enrichment has relevance to the
 cluster population in external galaxies.


\acknowledgements 
I would like express my appreciation to  David Lambert
  for letting me use some our
  collaborative work on RCBs. I also would like to acknowledge the help of many
  friends and
  collaborators who have supplied information or advice on various aspects,
  particularly Martin Asplund, Anibal Garcia-Hernandez, David Lambert,
  Gajendra Pandey, Betrand Plez, and Kjell Eriksson. I am most thankful to the
  editors of this proceddings Klaus Werner and Thomos Rauch for their patience
  and tolerance in condoning my undue delay in submitting this contribution.


\end{document}